\documentstyle[12pt,axodraw,graphicx]{article}
\begin{document}
\baselineskip 0.6cm

\def\simgt{\mathrel{\lower2.5pt\vbox{\lineskip=0pt\baselineskip=0pt
           \hbox{$>$}\hbox{$\sim$}}}}
\def\simlt{\mathrel{\lower2.5pt\vbox{\lineskip=0pt\baselineskip=0pt
           \hbox{$<$}\hbox{$\sim$}}}}

\begin{titlepage}

\begin{flushright}
UCB-PTH 04/07 \\
LBNL-54698 \\
\end{flushright}

\vskip 2.0cm

\begin{center}

{\Large \bf 
Matter Unification in Warped Supersymmetric SO(10)
}

\vskip 1.0cm

{\large
Yasunori Nomura$^{a,b}$ and David Tucker-Smith$^c$
}

\vskip 0.4cm

$^a$ {\it Department of Physics, University of California,
                Berkeley, CA 94720} \\
$^b$ {\it Theoretical Physics Group, Lawrence Berkeley National Laboratory,
                Berkeley, CA 94720} \\
$^c$ {\it Department of Physics, Williams College, 
                Williamstown, MA 01267}

\vskip 1.2cm

\abstract{We construct models of warped unification with a bulk $SO(10)$ 
gauge symmetry and boundary conditions that preserve the $SU(4)_C \times 
SU(2)_L \times SU(2)_R$ Pati-Salam gauge group (422).  In the dual 4D 
description, these models are 422 gauge theories in which the apparent 
unification of gauge couplings in the minimal supersymmetric standard 
model is explained as a consequence of strong coupling in the ultraviolet. 
The weakness of the gauge couplings at low energies is ensured in this 
4D picture by asymptotically non-free contributions from the conformal 
sector, which are universal due to an approximate $SO(10)$ global symmetry. 
The 422 gauge symmetry is broken to the standard model group by a simple 
set of Higgs fields.  An advantage of this setup relative to $SU(5)$ models 
of warped unification is that matter is automatically required to fill 
out representations of 422, providing an elegant understanding of the 
quantum numbers of the standard-model quarks and leptons.  The models 
also naturally incorporate the see-saw mechanism for neutrino masses 
and bottom-tau unification.  Finally, they predict a rich spectrum of 
exotic particles near the TeV scale, including states with different 
quantum numbers than those that appear in $SU(5)$ models.}

\end{center}
\end{titlepage}

\section{Introduction}

Two successful aspects of grand unification are the unification of 
gauge couplings and the unification of matter into a smaller number 
of representations.   Both features explain something about nature. 
The first explains why the running gauge couplings appear to meet at 
a very high energy in the context of weak scale supersymmetry.  The 
second helps to explain the quark and lepton gauge quantum numbers 
in the standard model.  The apparent unification of couplings can be 
easily addressed if the unified group is a simple group containing all 
the standard model gauge interactions.  The smallest successful group is 
$SU(5)$~\cite{Georgi:sy}, and the next smallest is $SO(10)$~\cite{SO10}. 
Of course, in these theories coupling unification and matter unification 
are closely related, as the enlarged gauge symmetry requires quarks and 
leptons to appear in representations of either $SU(5)$ or $SO(10)$. 

Based on matter unification alone, however, the simplest approach 
is arguably $SU(4)_C \times SU(2)_L \times SU(2)_R$ (422) {\it a la} 
Pati and Salam~\cite{Pati:1974yy}.  This group provides a very elegant 
explanation of the quantum numbers of the standard model fermions, 
with a full generation of quarks and leptons (including a right-handed 
neutrino) filling out the representations $({\bf 4}, {\bf 2}, {\bf 1}) 
+ ({\bf 4}^*, {\bf 1}, {\bf 2})$.  The breaking of the 422 group to 
its standard model subgroup is quite straightforward: it is attained 
by a simple set of Higgs fields, and this same set of fields is 
sufficient to break unwanted quark and lepton mass relations (while 
preserving bottom-tau unification) through non-renormalizable operators. 
This breaking also almost automatically leads to small neutrino masses 
through the see-saw mechanism (which is not the case in $SU(5)$). 
The simplicity of the gauge breaking makes 422 attractive compared 
with larger unified groups such as $SO(10)$, which require a more 
complicated structure for realistic gauge breaking. 

In this paper we explore the interesting role that the 422 gauge group 
can play in models of warped supersymmetric unification.  In the $SU(5)$ 
model introduced in~\cite{Goldberger:2002pc}, the unified gauge symmetry 
realized in the bulk is explicitly broken by boundary conditions on the 
Planck brane.  This implies that in the 4D dual description the theory 
does not have an $SU(5)$ gauge symmetry.  Rather, $SU(5)$ appears as an 
approximate {\em global} symmetry, possessed by the strongly interacting 
conformal sector that arises as the dual description of the bulk physics. 
The successful supersymmetric prediction relating the low-energy gauge 
couplings then follows from the assumption that the theory becomes 
strongly coupled in the ultraviolet.  The role of the $SU(5)$ global 
symmetry in this context is to ensure that the contributions from the 
conformal sector to the gauge coupling evolution are $SU(5)$ symmetric, 
so that the non-universal contributions are given purely by the elementary 
states, which are identical to the states of the minimal supersymmetric 
standard model (MSSM).  Thus, a simple group still plays an important 
role in the unification prediction, but in the infrared, rather than 
the ultraviolet, and as a global symmetry, rather than a gauge symmetry. 

This extra global symmetry does not require a complicated breaking 
mechanism -- it is explicitly broken from the beginning -- but nor does 
it require that the matter fields form unified representations.  Suppose 
we introduce quarks and leptons on the Planck brane in the $SU(5)$ model 
of~\cite{Goldberger:2002pc}.  Then there is no reason why they must fill out 
$SU(5)$ representations, and no reason why their hypercharges must satisfy 
the appropriate quantization condition.   This point becomes especially 
important in the class of models introduced in~\cite{NST}, where the unified 
symmetry is broken both on the Planck and TeV branes, in which case the quarks 
and leptons cannot arise from bulk matter alone.  We thus clearly need some 
other ingredient for understanding matter quantum numbers in these setups. 
We propose to use the 422 gauge group for this purpose.  We promote the 
gauge group in the 4D dual picture to 422 in these models.  This in turn 
requires that we promote the global group of the conformal sector, which 
is the bulk gauge group in the 5D picture, to $SO(10)$, at least.  The 
correct prediction relating the gauge couplings at low energies then 
arises through the global $SO(10)$ group, with the assumption of strong 
coupling.  Enlarging the global group to $SO(10)$ has direct consequences 
for physics at low energies, because it results in light exotics with 
different quantum numbers than those in the $SU(5)$ models.  The 
resulting phenomenology is, naturally, quite rich. 

The organization of the paper is as follows. In section~\ref{sec:model-1} 
we construct a model based on the structure of the model of~\cite{NST}, 
so that the bulk $SO(10)$ group is broken both on the Planck and TeV 
branes.  In section~\ref{sec:model-2} we present a model in which the 
bulk $SO(10)$ is broken only on the Planck brane, similarly to the model 
of~\cite{Goldberger:2002pc}.  In both theories the matter fields are 
located on the Planck brane, but without the problems such a setup 
produces in the $SU(5)$ models.  Moreover, we gain several desirable 
features coming from 422, including quark-lepton unification (and thus 
hypercharge quantization), bottom-tau unification, and a natural see-saw 
mechanism.  Some of the group theory in our models shares certain 
features with $SO(10)$ models in flat space~\cite{Asaka:2001eh}, but 
the physics involved is quite different.  Conclusions are given in 
section~\ref{sec:concl}.

\section{Model with TeV-Brane Symmetry Breaking}
\label{sec:model-1}

In this section we construct a model having the properties described 
in the introduction, and in which the bulk gauge symmetry is reduced 
at the TeV brane. This model can be viewed as an extension of the 
321-321 model of Ref.~\cite{NST}.  The main new ingredient of the 
model presented here is the unification of matter fields localized 
on the Planck brane. This leads to an understanding of the quark and 
lepton quantum numbers, as well as the ratio of the bottom-quark and 
tau-lepton masses through Yukawa unification.  Small neutrino masses 
also arise quite naturally through the see-saw mechanism.  All these 
features can be accommodated without spoiling the interesting features 
of the 321-321 model: automatic doublet-triplet splitting, suppression 
of proton decay, and a rich phenomenology of superparticles and 
grand-unified-theoretic (GUT) particles. 

\subsection{Basic setup}

The model is formulated in a 5D warped spacetime with the extra dimension 
$y$ compactified on an $S^1/Z_2$ orbifold: $0 \leq y \leq \pi R$.  The 
metric is given by 
\begin{equation}
  d s^2 = e^{-2k|y|} \eta_{\mu\nu} dx^\mu dx^\nu + dy^2,
\label{eq:metric}
\end{equation}
where $k$ is the AdS curvature, which is taken to be somewhat (typically 
a factor of a few) smaller than the 5D Planck scale $M_5$.  The 4D Planck 
scale, $M_{\rm Pl}$, is given by $M_{\rm Pl}^2 \simeq M_5^3/k$ and we 
take $k \sim M_5 \sim M_{\rm Pl}$.  We choose $kR \sim 10$ so that the 
TeV scale is naturally generated by the AdS warp factor: $k' \equiv 
k e^{-\pi kR} \sim {\rm TeV}$~\cite{Randall:1999ee}.

We choose the bulk gauge group to be $SO(10)$.  This bulk $SO(10)$ 
symmetry is then broken by boundary conditions imposed at the boundaries 
both at $y=0$ (Planck brane) and $\pi R$ (TeV brane).  Using the 4D 
$N=1$ superfield notation, in which the 5D gauge multiplet is described 
by a vector superfield $V(A_\mu, \lambda)$ and a chiral superfield 
$\Sigma(\sigma+iA_5, \lambda')$, the boundary conditions are given by 
\begin{equation}
  \pmatrix{V \cr \Sigma}(x^\mu,-y) 
  = \pmatrix{P V P^{-1} \cr -P \Sigma P^{-1}}(x^\mu,y), 
\qquad
  \pmatrix{V \cr \Sigma}(x^\mu,-y') 
  = \pmatrix{P V P^{-1} \cr -P \Sigma P^{-1}}(x^\mu,y'), 
\label{eq:bc-g-1}
\end{equation}
where $y' = y - \pi R$.  The matrix $P$ is chosen such that it leaves 
the Pati-Salam $SU(4)_C \times SU(2)_L \times SU(2)_R$ (422) subgroup of 
$SO(10)$ invariant. Specifically, in the basis where the generators of 
$SO(10)$, which are imaginary and antisymmetric $10 \times 10$ matrices, 
are given by $\sigma_0 \otimes A_5$, $\sigma_1 \otimes A_5$, $\sigma_2 
\otimes S_5$ and $\sigma_3 \otimes A_5$, the matrix $P$ can be chosen as 
$P = \sigma_0 \otimes {\rm diag}(1,1,1,-1,-1)$. Here, $\sigma_0$ is the 
$2 \times 2$ unit matrix and $\sigma_{1,2,3}$ are the Pauli spin matrices; 
$S_5$ and $A_5$ are $5 \times 5$ matrices that are real and symmetric, 
and imaginary and antisymmetric, respectively.  This reduces the gauge 
group at the Planck and the TeV branes to 422, and leaves the $({\bf 15}, 
{\bf 1}, {\bf 1}) + ({\bf 1}, {\bf 3}, {\bf 1}) + ({\bf 1}, {\bf 1}, 
{\bf 3})$ component of $V$ and the $({\bf 6}, {\bf 2}, {\bf 2})$ 
component of $\Sigma$ as zero modes, where the numbers in parentheses 
represent quantum numbers under 422.  All components of the $SO(10)$ 
gauge multiplet have Kaluza-Klein (KK) towers with the typical mass 
scale of $k' \sim {\rm TeV}$. 

The Higgs fields are introduced in the bulk as a hypermultiplet 
transforming as ${\bf 10}$ of $SO(10)$, which is described by two chiral 
superfields as $\{ H, H^c \}$ in the 4D $N=1$ superfield notation.  They 
obey the boundary conditions 
\begin{equation}
  \pmatrix{H \cr H^c}(x^\mu,-y) 
  = \pmatrix{-P H \cr P H^c}(x^\mu,y), 
\qquad
  \pmatrix{H \cr H^c}(x^\mu,-y') 
  = \pmatrix{-P H \cr P H^c}(x^\mu,y'), 
\label{eq:bc-h-1}
\end{equation}
which leave the $({\bf 1}, {\bf 2}, {\bf 2})$ component of $H$ and 
the $({\bf 6}, {\bf 1}, {\bf 1})$ component of $H^c$ as zero modes.%
\footnote{Alternatively, we could introduce the Higgs field $H$ on the 
Planck brane in the $({\bf 1}, {\bf 2}, {\bf 2})$ representation of 422. 
Another possibility is to put the Higgs fields in the bulk and impose 
the boundary conditions of Eq.~(\ref{eq:bc-h-1}) with an extra minus 
sign in the right-hand side of the second equation (i.e.~flipping the 
TeV-brane boundary conditions)~\cite{NST}. \label{ft:Higgs}} 
The Higgs multiplet can have a mass parameter in the bulk, which we 
parameterize as $c_H^{} k$.  The parameter $c_H$ then controls the 
wavefunction profiles for the zero modes arising from $\{ H, H^c \}$. 
As in the 321-321 model of~\cite{NST}, the unwanted zero modes from 
$\Sigma$ and $H^c$ obtain masses when supersymmetry is broken. 

Matter fields are introduced on the Planck brane as chiral superfields 
in the $\Psi({\bf 4}, {\bf 2}, {\bf 1}) + \bar{\Psi}({\bf 4}^*, {\bf 1}, 
{\bf 2})$ representation of 422 for each generation, which  contain our 
quarks and leptons (and a right-handed neutrino) as $\Psi = \{ Q, L \}$ 
and $\bar{\Psi} = \{ U, D, E, N \}$.  Since the gauge group on the Planck 
brane is non-Abelian, the charges of the matter fields are quantized. 
This setup also requires the existence of right-handed neutrinos, which 
is an important ingredient for the see-saw mechanism. 

To reproduce successful phenomenology at low energies, the low-energy 
gauge group must be reduced to $SU(3)_C \times SU(2)_L \times U(1)_Y$ 
(321). To this end, we break the 422 group by the Higgs mechanism on 
the Planck brane.  The simplest possibility is to introduce chiral 
superfields on the Planck brane in the $\chi({\bf 4}, {\bf 1}, {\bf 2}) 
+ \bar{\chi}({\bf 4}^*, {\bf 1}, {\bf 2})$ representation of 422, 
and give appropriate vacuum expectation values for them.%
\footnote{We could also introduce $\chi'({\bf 4}, {\bf 2}, {\bf 1}) 
+ \bar{\chi}'({\bf 4}^*, {\bf 2}, {\bf 1})$ fields on the brane and 
make the breaking of left-right symmetry entirely spontaneous.}
The expectation values are easily induced, for example, by introducing 
the superpotential interaction $\int\!d^2\theta S(\chi \bar{\chi} - 
v_\chi^2) + {\rm h.c.}$ on the Planck brane, where $S$ is a singlet 
chiral superfield.  Here we take the expectation values $\langle \chi 
\rangle = \langle \bar{\chi} \rangle = v_\chi$ to be of order $k$, 
which is a natural scale on the Planck brane.  This then breaks the 
gauge group on the Planck brane to 321 at the scale $k$, and gives masses 
to the $422/321$ component of $V$, which would otherwise be massless. 

Quark and lepton masses are generated on the Planck brane through 
the following operators: 
\begin{equation}
  S = \int\!d^4x \int_0^{\pi R}\!\!dy \,\, 
    2 \delta(y) \biggl[ \int\!d^2\theta \left( y\, \Psi \bar{\Psi} H_D 
    + \frac{\lambda}{M_*} (\chi \bar{\Psi})^2 \right)
    + {\rm h.c.} \biggr],
\label{eq:yukawa-422}
\end{equation}
where $H_D$ represents the $({\bf 1}, {\bf 2}, {\bf 2})$ component 
of $H$ under the 422 decomposition, and we have omitted generation 
indices. $M_*$ is the cutoff scale of order $M_5$.  The first term 
gives Yukawa couplings for quarks and leptons while the second term 
gives Majorana masses for right-handed neutrinos of order $\langle 
\chi \rangle^2/M_*$, which generates small masses for the observed 
neutrinos through the see-saw mechanism.  The unwanted mass relations 
arising from Eq.~(\ref{eq:yukawa-422}) for the first-two generation 
quarks and leptons are broken by higher dimensional operators involving 
$\langle \chi \rangle$ and $\langle \bar{\chi} \rangle$, allowing for 
realistic quark and lepton masses and mixings.%
\footnote{The higher dimensional operators that allow for realistic 
fermion masses have the same form as those employed in 4D 422 theories, 
see e.g.~\cite{King:fv}.}
For relatively suppressed 422-breaking expectation values $\langle \chi 
\rangle = \langle \bar{\chi} \rangle \approx k \simlt M_*$, the Yukawa 
couplings for the third generation fermions are approximately unified at the 
scale $k$, which gives a successful $m_b/m_\tau$ prediction at low energies. 
It also leads the theory to the large $\tan\beta$ region, $\tan\beta \equiv 
\langle H_u \rangle/\langle H_d \rangle \approx 50$, where $H_u$ and $H_d$ 
($\subset H_D$) are the Higgs fields giving masses to the up-type and 
down-type quarks, respectively.  The couplings in Eq.~(\ref{eq:yukawa-422}) 
respect a $U(1)_R$ symmetry with the charges given by $V(0), \Sigma(0), H(0), 
H^c(2), \Psi(1), \bar{\Psi}(1), \chi(0), \bar{\chi}(0)$.  This symmetry, 
when imposed on the theory, forbids potentially dangerous operators, 
such as $\int\!d^2\theta H_D^2 + {\rm h.c.}$, on the Planck brane. As in 
the $SU(5)$ models, proton decay is not a problem for the Planck-brane 
localized matter because all potentially dangerous gauge bosons (and 
their KK towers) have wavefunctions strongly peaked towards the TeV brane. 

\subsection{Prediction for gauge couplings}

\begin{figure}
\begin{center}
\begin{picture}(400,520)(-45,-25)
%
%
 \Line(-2,350)(325,350) \Text(-30,500)[r]{\Large (a)}
 \LongArrow(-2,350)(-2,500) \Text(-5,500)[r]{$m_n$} \Text(-7,350)[r]{0}
   \Line(5,350)(25,350) \Vertex(15,350){2} 
   \Line(5,433)(25,433) \Vertex(15,433){2}
   \Line(30,350)(50,350) \Vertex(40,350){2}
   \Line(30,433)(50,433) \Vertex(35,433){2} \Vertex(45,433){2}
   \Line(55,433)(75,433) \Vertex(65,433){2}
   \Line(80,433)(100,433) \Vertex(90,433){2}
   \Line(115,433)(135,433) \Vertex(125,433){2}
   \Line(140,433)(160,433) \Vertex(145,433){2} \Vertex(155,433){2}
   \Line(165,433)(185,433) \Vertex(175,433){2}
   \Line(190,433)(210,433) \Vertex(200,433){2}
   \Line(225,479)(245,479) \Vertex(235,479){2}
   \Line(250,350)(270,350) \Vertex(260,350){2}
   \Line(250,479)(270,479) \Vertex(255,479){2} \Vertex(265,479){2}
   \Line(275,350)(295,350) \Vertex(285,350){2}
   \Line(275,479)(295,479) \Vertex(285,479){2}
   \Line(300,350)(320,350) \Vertex(310,350){2}
   \Line(300,479)(320,479) \Vertex(310,479){2}
%
%
 \Line(-2,175)(325,175) \Text(-30,325)[r]{\Large (b)}
 \LongArrow(-2,175)(-2,325) \Text(-5,325)[r]{$m_n$} \Text(-7,175)[r]{0}
   \Line(5,175)(25,175) \Vertex(15,175){2} 
   \Line(5,258)(25,258) \Vertex(15,258){2}
   \Line(30,178)(50,178) \Vertex(40,178){2}
   \Line(30,240)(50,240) \Vertex(40,240){2} \LongArrow(40,256)(40,243)
   \Line(30,268)(50,268) \Vertex(40,268){2} \LongArrow(40,260)(40,265)
   \Line(55,258)(75,258) \Vertex(65,258){2}
   \Line(80,258)(100,258) \Vertex(90,258){2}
 \DashLine(25,258)(55,258){2}
   \Line(115,258)(135,258) \Vertex(125,258){2}
   \Line(140,240)(160,240) \Vertex(150,240){2} \LongArrow(150,256)(150,243)
   \Line(140,268)(160,268) \Vertex(150,268){2} \LongArrow(150,260)(150,265)
   \Line(165,258)(185,258) \Vertex(175,258){2}
   \Line(190,258)(210,258) \Vertex(200,258){2}
 \DashLine(135,258)(165,258){2}
   \Line(225,304)(245,304) \Vertex(235,304){2}
   \Line(250,198)(270,198) \Vertex(260,198){2} \LongArrow(260,177)(260,195) 
   \Line(250,293)(270,293) \Vertex(260,293){2} \LongArrow(260,302)(260,296)
   \Line(250,318)(270,318) \Vertex(260,318){2} \LongArrow(260,306)(260,315)
   \Line(275,202)(295,202) \Vertex(285,202){2} \LongArrow(285,177)(285,199) 
   \Line(275,308)(295,308) \Vertex(285,308){2}
   \Line(300,181)(320,181) \Vertex(310,181){2}
   \Line(300,304)(320,304) \Vertex(310,304){2}
 \DashLine(245,304)(300,304){2}
%
%
 \Line(-2,0)(325,0) \Text(-30,150)[r]{\Large (c)}
 \LongArrow(-2,0)(-2,150) \Text(-5,150)[r]{$m_n$} \Text(-7,0)[r]{0}
 \Text(15,-19)[b]{$A_\mu^{321}$}
   \Line(5,0)(25,0) \Vertex(15,0){2} 
   \Line(5,83)(25,83) \Vertex(15,83){2}
 \Text(40,-19)[b]{$\lambda^{321}$}
   \Line(30,8)(50,8) \Vertex(35,8){2} \LongArrow(40,0)(40,7)
   \Line(30,10)(50,10) \Vertex(45,10){2} \LongArrow(40,81)(40,13)
   \Line(30,128)(50,128) \Vertex(35,128){2} \LongArrow(40,85)(40,125)
   \Line(30,131)(50,131) \Vertex(45,131){2} \LongArrow(40,150)(40,134)
 \Text(65,-19)[b]{$\sigma^{321}$}
   \Line(55,83)(75,83) \Vertex(65,83){2}
 \Text(90,-19)[b]{$A_5^{321}$}
   \Line(80,83)(100,83) \Vertex(90,83){2}
 \DashLine(25,83)(55,83){2}
 \Text(125,-19)[b]{$A_\mu^{\rm PS}$}
   \Line(115,83)(135,83) \Vertex(125,83){2}
 \Text(150,-19)[b]{$\lambda^{\rm PS}$}
   \Line(140,5)(160,5) \Vertex(150,5){2} \LongArrow(150,81)(150,8)
   \Line(140,128)(160,128) \Vertex(145,128){2} \LongArrow(150,85)(150,125)
   \Line(140,131)(160,131) \Vertex(155,131){2} \LongArrow(150,150)(150,134)
 \Text(175,-19)[b]{$\sigma^{\rm PS}$}
   \Line(165,83)(185,83) \Vertex(175,83){2}
 \Text(200,-19)[b]{$A_5^{\rm PS}$}
   \Line(190,83)(210,83) \Vertex(200,83){2}
 \DashLine(135,83)(165,83){2}
 \Text(235,-19)[b]{$A_\mu^{\rm XY}$}
   \Line(225,129)(245,129) \Vertex(235,129){2}
 \Text(260,-19)[b]{$\lambda^{\rm XY}$}
   \Line(250,81)(270,81) \Vertex(255,81){2} \LongArrow(260,2)(260,78) 
   \Line(250,84)(270,84) \Vertex(265,84){2} \LongArrow(260,127)(260,87)
   \Line(260,131)(260,150)
 \Text(285,-19)[b]{$\sigma^{\rm XY}$}
   \Line(275,82)(295,82) \Vertex(285,82){2} \LongArrow(285,2)(285,79) 
   \Line(285,131)(285,150)
 \Text(310,-19)[b]{$A_5^{\rm XY}$}
   \Line(300,12)(320,12) \Vertex(310,12){2} \LongArrow(310,1)(310,9)
   \Line(300,129)(320,129) \Vertex(310,129){2}
 \DashLine(245,129)(300,129){2}
\end{picture}
\caption{Schematic depiction for the lowest-lying masses for the 
 gauge multiplet.  The three figures represent the spectrum (a) in the 
 supersymmetric limit, (b) for small supersymmetry breaking and (c) 
 for large supersymmetry breaking.  Each bullet for $\lambda^{321}$ 
 ($\lambda^{\rm PS}$ and $\lambda^{\rm XY}$) represents a Majorana 
 (Dirac) degree of freedom.}
\label{fig:spectrum}
\end{center}
\end{figure}
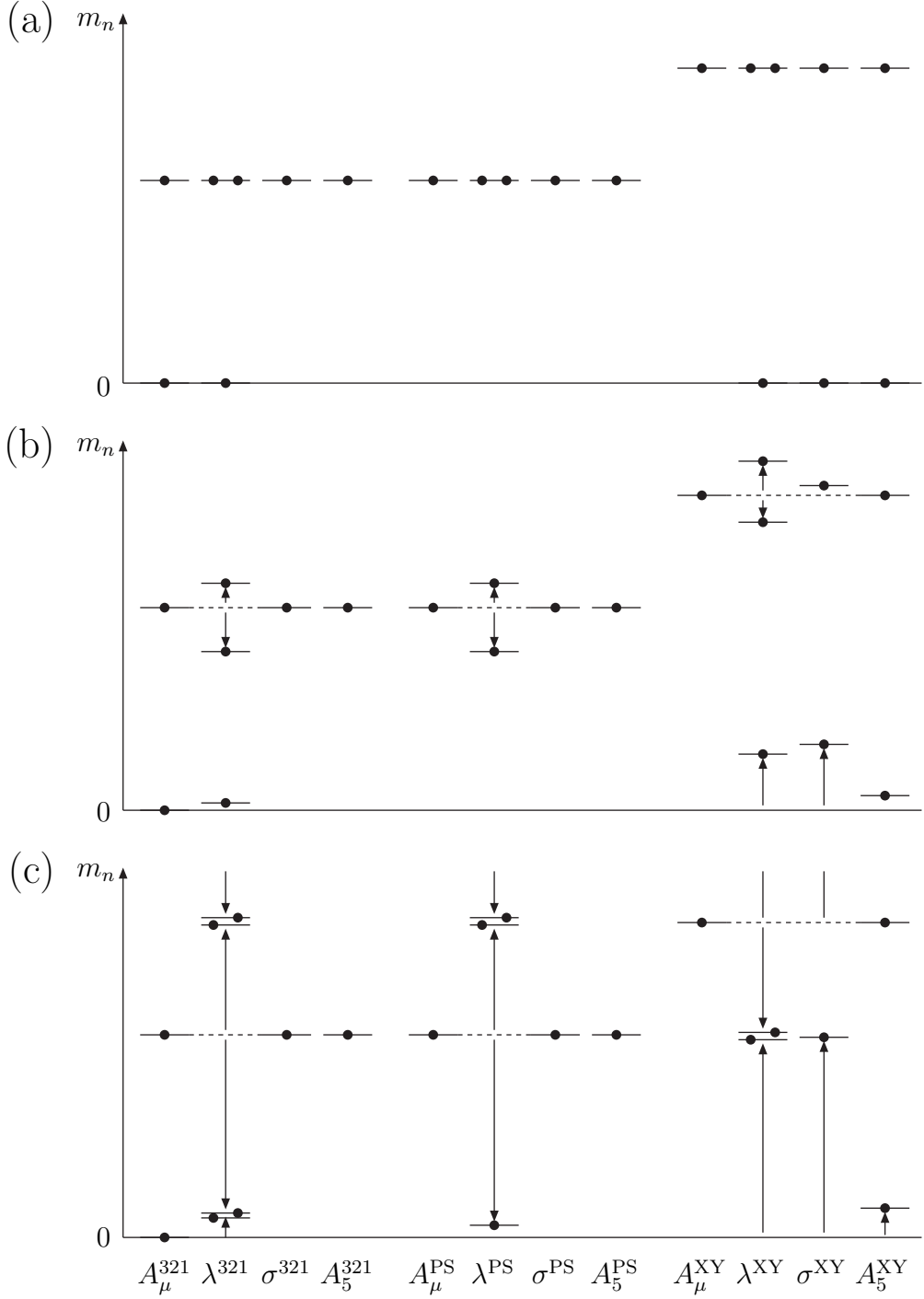

The $SO(10)$ generators in the present model are naturally divided into 
three classes: (i) the 321 generators, (ii) the generators belonging 
to 422/321, which we call PS, and (iii) the generators belonging to 
$SO(10)$/422, which we call XY.  The spectrum of the gauge sector is 
then given as follows. The 321 gauge multiplet has a zero mode $V^{321}$ 
and a KK tower, which consists of $V^{321}$ and $\Sigma^{321}$ at each 
KK level, with the masses $m_n$ given by the solutions of 
\begin{equation}
  \frac{J_0\left(\frac{m_n}{k}\right)
    + \frac{g_B^2}{\tilde{g}_{0,a}^2}m_n J_1\left(\frac{m_n}{k}\right)}
  {Y_0\left(\frac{m_n}{k}\right)
    + \frac{g_B^2}{\tilde{g}_{0,a}^2}m_n Y_1\left(\frac{m_n}{k}\right)}
  = \frac{J_0\left(\frac{m_n}{k'}\right)}
    {Y_0\left(\frac{m_n}{k'}\right)},
\label{eq:KKmass-321}
\end{equation}
where $J_n(x)$ and $Y_n(x)$ are the Bessel functions of order $n$, and 
$a=1,2,3$ represents $U(1)_Y$, $SU(2)_L$ and $SU(3)_C$, respectively; 
$g_B$ is the bulk $SO(10)$ gauge coupling and $\tilde{g}_{0,a}^2$ are 
the Planck-brane gauge couplings appropriately renormalized at the scale 
$k'$ (for more precise definitions and our assumptions regarding the 
ultraviolet values of these parameters, see~\cite{NST}). The PS gauge 
multiplet does not have a zero-mass mode, and its KK tower consists 
of $V^{\rm PS}$ and $\Sigma^{\rm PS}$ at each KK level with the 
masses approximately given by%
\footnote{More precisely, the left-hand side of Eq.~(\ref{eq:KKmass-PS}) 
is given by $\{J_0(m_n/k)-(C g_B^2 v_\chi^2/m_n)J_1(m_n/k)\}/%
\{Y_0(m_n/k)-(C g_B^2 v_\chi^2/m_n)Y_1(m_n/k)\}$, where $C$ is an $O(1)$ 
coefficient depending on the gauge component.  For $v_\chi \gg k'$, 
however, the denominator is well approximated by $\approx -(C g_B^2 
v_\chi^2/m_n)Y_1(m_n/k)$, which is enough to guarantee that the mass 
eigenvalues are almost given by the zeros of the right-hand side of 
Eq.~(\ref{eq:KKmass-PS}).  The expression for the left-hand side further 
simplifies for $v_\chi \approx k$ and $g_B^2 k \gg 1$ to $J_1(m_n/k)/%
Y_1(m_n/k)$, which we have used in Eq.~(\ref{eq:KKmass-PS}). However, 
this is not essential because the solutions are quite insensitive to 
the detailed expression of the left-hand side for $v_\chi \gg k'$. 
The same comment applies also to the expressions involving the PS 
gauginos, e.g. Eq.~(\ref{eq:KKmass-PSgauginos}). \label{ft:comment}}
\begin{equation}
  \frac{J_1\left(\frac{m_n}{k}\right)}{Y_1\left(\frac{m_n}{k}\right)}
  = \frac{J_0\left(\frac{m_n}{k'}\right)}{Y_0\left(\frac{m_n}{k'}\right)}.
\label{eq:KKmass-PS}
\end{equation}
Finally, the XY gauge multiplet has a zero mode $\Sigma^{\rm XY}$, and 
its KK tower $\{ V^{\rm XY}, \Sigma^{\rm XY} \}$ has masses given by 
\begin{equation}
  \frac{J_1\left(\frac{m_n}{k}\right)}
    {Y_1\left(\frac{m_n}{k}\right)}
  = \frac{J_1\left(\frac{m_n}{k'}\right)}
    {Y_1\left(\frac{m_n}{k'}\right)}.
\label{eq:KKmass-XY}
\end{equation}
This spectrum can be summarized, for $k' \ll k$, as 
\begin{equation}
  \left\{ \begin{array}{ll} 
    V^{321},\, \Sigma^{\rm XY} : & m_0 = 0, \\
    \{ V^{321}, \Sigma^{321} \} + \{ V^{\rm PS}, \Sigma^{\rm PS} \}: 
      & m_n \simeq (n-\frac{1}{4})\pi k', \\
    \{ V^{\rm XY}, \Sigma^{\rm XY} \}: 
      & m_n \simeq (n+\frac{1}{4})\pi k',
  \end{array} \right.
\label{eq:spectrum-SO10-1}
\end{equation}
where $n = 1,2,\cdots$.  The transformation properties of these fields 
under the 321 gauge group are given by $({\bf 8}, {\bf 1})_0 + ({\bf 1}, 
{\bf 3})_0 + ({\bf 1}, {\bf 1})_0$ for $V^{321}$ and $\Sigma^{321}$, 
$({\bf 3}, {\bf 1})_{2/3} + ({\bf 3}^*, {\bf 1})_{-2/3} + ({\bf 1}, 
{\bf 1})_1 + ({\bf 1}, {\bf 1})_{-1} + ({\bf 1}, {\bf 1})_0$ for 
$V^{\rm PS}$ and $\Sigma^{\rm PS}$, and $({\bf 3}, {\bf 2})_{-5/6} 
+ ({\bf 3}^*, {\bf 2})_{5/6} + ({\bf 3}, {\bf 2})_{1/6} + ({\bf 3}^*, 
{\bf 2})_{-1/6}$ for $V^{\rm XY}$ and $\Sigma^{\rm XY}$.  A schematic 
depiction of the spectrum is given in Fig.~\ref{fig:spectrum}a.

With the gauge spectrum of Eq.~(\ref{eq:spectrum-SO10-1}) and the Higgs 
fields of Eq.~(\ref{eq:bc-h-1}), the MSSM prediction for the low-energy gauge 
couplings is preserved.  Specifically, the low-energy 321 gauge couplings are 
given by setting $(T_1,T_2,T_3)(V_{++}) = (0,2,3)$, $(T_1,T_2,T_3)(V_{+-}) = 
(0,0,0)$, $(T_1,T_2,T_3)(V_{-+}) = (14/5,0,1)$ and $(T_1,T_2,T_3)(V_{--}) = 
(26/5,6,4)$ in Eq.~(9) of Ref.~\cite{Goldberger:2002pc} and adding the Higgs 
contribution, which is identical to that of the 321-321 model (the spectrum 
of the PS gauge multiplet is effectively reproduced by imposing the $(-,+)$ 
and $(+,-)$ boundary conditions on $V^{\rm PS}$ and $\Sigma^{\rm PS}$, 
respectively, and the Higgs spectrum is identical to the $SU(5)$ Higgs 
spectrum of~\cite{NST} with $c_H = c_{H({\bf 5})} = c_{\bar{H}({\bf 5}^*)}$). 
We then find that for $c_H \geq 1/2$ the prediction for low-energy 321 
gauge couplings $g_a$ is given by 
\begin{equation}
  \frac{1}{g_a^2(k')} 
  \simeq ({\rm universal}) 
    + \frac{1}{8 \pi^2} \Delta^a,
\label{eq:gc-low-1}
\end{equation}
where
\begin{equation}
  \pmatrix{\Delta^1 \cr \Delta^2 \cr \Delta^3}
    \simeq \pmatrix{33/5 \cr 1 \cr -3} \ln\left(\frac{k}{k'}\right),
\label{eq:gc-low-2}
\end{equation}
hence reproducing the successful MSSM prediction.%
\footnote{Successful gauge coupling unification in warped unified 
theories was anticipated in~\cite{Pomarol:2000hp} based on a heuristic 
argument, and was shown explicitly in~\cite{Goldberger:2002pc}. 
Techniques for calculating gauge coupling evolution in warped 
space were developed in Refs.~\cite{Randall:2001gc}.}
In a suitable renormalization scheme, the logarithmic contribution 
of Eq.~(\ref{eq:gc-low-2}) arises entirely from the Planck 
brane couplings $\tilde{g}_{0,a}$, implying that we should use 
$1/\tilde{g}_{0,a}^2 \simeq \Delta^a/8\pi^2$ in Eq.~(\ref{eq:KKmass-321}). 
The result of Eqs.~(\ref{eq:gc-low-1},~\ref{eq:gc-low-2}) can also be 
understood in the 4D dual picture as follows.  In the 4D picture the 
theory between $v_\chi \approx k$ and the TeV scale is described by 
an $N=1$ supersymmetric gauge theory with the gauge group $SU(3)_C 
\times SU(2)_L \times U(1)_Y \times G$ with the $G$ sector possessing 
a global $SO(10)$ symmetry, where $G$ represents some gauge interaction 
whose coupling evolves very slowly over this energy interval.  The 
quark, lepton and Higgs doublets are interpreted as elementary fields, 
while various GUT states are regarded as composite states arising from 
the non-trivial infrared (of order TeV) dynamics of $G$. Since the 
contribution from the $G$ sector is universal due to the global $SO(10)$, 
the differences among the low-energy 321 gauge couplings arise entirely 
from the contribution of the elementary fields (assuming strong coupling 
at ultraviolet, see~\cite{Goldberger:2002pc}).  This gives the desired 
MSSM prediction because the elementary sector is identical to the MSSM. 

Here we comment on calculability in this model.  In a theory of 
warped supersymmetric unification, the size of the bulk gauge coupling 
$g_B$ is related to the 4D gauge coupling as $1/g_4^2 = \pi R/g_B^2$, 
where $g_4$ represents the unified gauge coupling in conventional 4D 
supersymmetric unification, $g_4 \simeq 0.7$.  Defining $M_*$ to 
be the scale where the 5D theory becomes strongly coupled, we obtain 
$1/g_B^2 \simeq C M_*/L$, where $C$ and $L$ are the group-theoretic 
and 5D-loop factors, respectively~\cite{Hall:2001xb}.  For the 
$SO(10)$ theory, $C \simeq 8$.  Using the 5D-loop factor of $L \simeq 
24\pi^3$~\cite{Chacko:1999hg} and $kR \sim 10$, we obtain $M_*/\pi k 
\simeq 2$.  This implies that the infrared cutoff of the theory, 
$M'_* \equiv M_* e^{-\pi kR}$, is close to the scale of the first KK 
excitation, $\pi k'$: $M'_*/\pi k' \simeq 2$.  This strongly restricts 
calculability --- we generically expect errors of order $(\pi k'/M'_*)^n$ 
in various predictions, where $n$ depends on the quantity (errors for 
the masses of the lightest 321 gauginos could be suppressed further 
by $1/\ln(k/k')$).%
\footnote{The situation is somewhat better in theories with the bulk 
$SU(5)$ symmetry because of the smaller value of $C$: $C \simeq 5$.}
The equations that follow should thus be interpreted with care: their 
precision is not very high and the results for higher KK towers are 
not meaningful. In general, this is the case for any warped theory 
with a large bulk gauge symmetry, so the same comment also applies 
to the model presented in the next section.  We stress, however, that 
the main features of the model, such as its unified understanding of 
matter quantum numbers and the qualitative aspects of its spectrum, 
are not affected by these numerical limitations. 

\subsection{Supersymmetry breaking}

We now consider the effects of supersymmetry breaking in the present 
model.  Here we follow the notation of Ref.~\cite{NST}. Supersymmetry 
breaking is introduced on the TeV brane through the following 
potential~\cite{Gherghetta:2000qt}:
\begin{equation}
  S = \int\!d^4x \int_0^{\pi R}\!\!dy \,\, 
    2 \delta(y-\pi R) \biggl[ e^{-2\pi kR}\! \int\!d^4\theta Z^\dagger Z
    + \biggl\{ e^{-3\pi kR}\! \int\!d^2\theta \Lambda^2 Z + 
      {\rm h.c.} \biggr\} \biggr],
\label{eq:Z-TeV}
\end{equation}
where $Z$ is a singlet chiral superfield and $\Lambda$ is a mass parameter 
of order $M_* \sim M_5$.  This potential gives the vacuum expectation 
value $\langle Z \rangle = - e^{-\pi kR} \Lambda^{*2} \theta^2$, breaking 
supersymmetry and the $U(1)_R$ symmetry (to the $Z_{2,R}$ subgroup). This 
breaking does not destroy the successful prediction relating the low-energy 
gauge couplings, although it causes a distortion of the spectrum.

The masses for the 321 and PS gauginos, $\lambda^{321}$ and 
$\lambda^{\rm PS}$, are generated through the operators on the TeV brane 
of the form $\int\!d^2\theta Z {\rm Tr}[{\cal W}^\alpha {\cal W}_{\alpha}]
+ {\rm h.c.}$\, Since the gauge symmetry on the TeV brane is 422, we have 
three independent coefficients $\zeta_C$, $\zeta_L$ and $\zeta_R$ for 
these operators, corresponding to the $SU(4)_C$, $SU(2)_L$ and $SU(2)_R$ 
factors, respectively. Specifically, the operators are given by
\begin{equation}
  S = \int\!d^4x \int_0^{\pi R}\!\!dy\,
    2 \delta(y-\pi R) \!\!\sum_{A=C,L,R} \biggl[ 
      -\int\!d^2\theta\, \frac{\zeta_A}{2 M_*} Z \, 
      {\rm Tr}[ {\cal W}_A^\alpha {\cal W}_{A \alpha} ] 
      + {\rm h.c.} \biggr],
\label{eq:gaugino-mass}
\end{equation}
where $A=C,L,R$ denotes $SU(4)_C$, $SU(2)_L$ and $SU(2)_R$.  This implies 
that the masses for the 321 gauginos $\lambda^{321}_a$ ($a=1,2,3$) 
and the PS gauginos $\lambda^{\rm PS}_U$, $\lambda^{\rm PS}_E$ and 
$\lambda^{\rm PS}_S$ are all determined in terms of four parameters 
$\zeta_C M'$, $\zeta_L M'$, $\zeta_R M'$ and $M'/k'$, where $M' 
\equiv e^{-\pi kR} \Lambda^{*2}/M_*$.  Here $\lambda^{\rm PS}_U$, 
$\lambda^{\rm PS}_E$, and $\lambda^{\rm PS}_S$ transform under 321 
as $({\bf 3}, {\bf 1})_{2/3} + ({\bf 3}^*, {\bf 1})_{-2/3}$, $({\bf 1}, 
{\bf 1})_1 + ({\bf 1}, {\bf 1})_{-1}$, and $({\bf 1}, {\bf 1})_0$, 
respectively.  Moreover, if left-right symmetry is unbroken on the TeV brane, 
which is natural if left-right symmetry is broken spontaneously only on the 
Planck brane, we have an additional relation $\zeta_L = \zeta_R$. In this 
case the masses for the above gauginos are all determined by the three 
parameters $\zeta_C M'$, $\zeta_L M'$ and $M'/k'$.

Solving the equations of motion in 5D, we find that the masses for 
the $SU(3)_C$ and $SU(2)_L$ gauginos ($a=3$ and $2$ respectively) are 
given by 
\begin{equation}
  \frac{J_0\left(\frac{m_n}{k}\right)
    + \frac{g_B^2}{\tilde{g}_{0,a}^2}m_n J_1\left(\frac{m_n}{k}\right)}
  {Y_0\left(\frac{m_n}{k}\right)
    + \frac{g_B^2}{\tilde{g}_{0,a}^2}m_n Y_1\left(\frac{m_n}{k}\right)}
  = \frac{J_0\left(\frac{m_n}{k'}\right)
    + g_B^2 M_{\lambda,a} J_1\left(\frac{m_n}{k'}\right)}
    {Y_0\left(\frac{m_n}{k'}\right)
    + g_B^2 M_{\lambda,a} Y_1\left(\frac{m_n}{k'}\right)},
\label{eq:KKmass-32gauginos}
\end{equation}
where $M_{\lambda,3} = \zeta_C \Lambda^{*2}/M_*$ and $M_{\lambda,2} 
= \zeta_L \Lambda^{*2}/M_*$.  The masses for the $SU(3)_C$ and $SU(2)_L$ 
gaugino towers are given as the solutions to this equation, which can 
be $m_n < 0$ as well as $m_n > 0$ (the physical masses are given by 
$|m_n|$).  The masses for the PS gauginos $\lambda^{\rm PS}_U$ and 
$\lambda^{\rm PS}_E$ are similarly given by 
\begin{equation}
  \frac{J_1\left(\frac{m_n}{k}\right)}{Y_1\left(\frac{m_n}{k}\right)}
  = \frac{J_0\left(\frac{m_n}{k'}\right)
    - g_B^2 M_{\lambda,A} J_1\left(\frac{m_n}{k'}\right)}
    {Y_0\left(\frac{m_n}{k'}\right)
    - g_B^2 M_{\lambda,A} Y_1\left(\frac{m_n}{k'}\right)},
\label{eq:KKmass-PSgauginos}
\end{equation}
where $A$ takes $U$ and $E$ for $\lambda^{\rm PS}_U$ and 
$\lambda^{\rm PS}_E$ respectively, and $M_{\lambda,U} = \zeta_C 
\Lambda^{*2}/M_*$ and $M_{\lambda,E} = \zeta_R \Lambda^{*2}/M_*$. 
An interesting point is that for large supersymmetry breaking (i.e. 
large $\Lambda$) the lowest $\lambda^{321}_3$ and $\lambda^{321}_2$ 
gauginos both become pseudo-Dirac states with the masses $\approx 
(2/\pi kR)^{1/2}k' \simeq k'/4$, while the lowest $\lambda^{\rm PS}_U$ 
and $\lambda^{\rm PS}_E$ modes become very light with the masses given by 
$\approx (2/g_B^2 M_{\lambda,U})k'$ and $\approx (2/g_B^2 M_{\lambda,E})k'$, 
respectively.  This can be easily understood by noticing that the form 
of Eq.~(\ref{eq:KKmass-32gauginos}) and Eq.~(\ref{eq:KKmass-PSgauginos}) 
are identical, respectively, to the equations determining the masses of 
the 321 and $SU(5)/321$ gauginos in the model of~\cite{Goldberger:2002pc,%
Nomura:2003qb}. For small supersymmetry breaking, the lowest modes 
of $\lambda^{321}_3$ and $\lambda^{321}_2$ (i.e. the MSSM gauginos) 
have masses much smaller than $k'$, while $\lambda^{\rm PS}_U$ 
and $\lambda^{\rm PS}_E$ do not have a mode lighter than $k'$. 

The masses for the $U(1)_Y$ gaugino, $\lambda^{321}_1$, and the 
PS gaugino $\lambda^{\rm PS}_S$ obey a somewhat more complicated 
equation, since they generically mix with each other. The gauginos 
$\lambda^{321}_1$ and $\lambda^{\rm PS}_S$ are associated with two $U(1)$ 
factors arising from $SU(4)_C \times SU(2)_R$: $U(1)_Y \subset 321$ and 
$U(1)_\chi$, respectively.  Taking the two $U(1)$'s to be orthogonal, 
$U(1)_\chi$ is the ``fiveness'' charge arising as $U(1)_\chi \subset 
SO(10)/SU(5)$ in the standard GUT embedding.  The generators for $U(1)_Y$ 
and $U(1)_\chi$, $Y$ and $Q_S$, normalized in the $SO(10)$ covariant 
manner are given by $Y = \sqrt{2/5}\,T^C_{15} + \sqrt{3/5}\,T^R_{3}$ and 
$Q_S = -\sqrt{3/5}\,T^C_{15} + \sqrt{2/5}\,T^R_{3}$, where $T^C_{15}$ 
($T^R_{3}$) is a generator of $SU(4)_C$ ($SU(2)_R$) that commutes with 
321. The masses for these gauginos are then given by the equation
\begin{eqnarray}
  && \Biggl( J_Y\Bigl(\frac{m_n}{k'}\Bigl) 
    - \frac{\tilde{J}_0\bigl(\frac{m_n}{k}\bigl)}
      {\tilde{Y}_0\bigl(\frac{m_n}{k}\bigl)}
      Y_Y\Bigl(\frac{m_n}{k'}\Bigl) \Biggr)
  \Biggl( J_S\Bigl(\frac{m_n}{k'}\Bigl) 
    - \frac{J_1\bigl(\frac{m_n}{k}\bigl)}{Y_1\bigl(\frac{m_n}{k}\bigl)}
      Y_S\Bigl(\frac{m_n}{k'}\Bigl) \Biggr)
\nonumber\\
  && \qquad \qquad - 
  \Biggl( J_M\Bigl(\frac{m_n}{k'}\Bigl) 
    - \frac{\tilde{J}_0\bigl(\frac{m_n}{k}\bigl)}
      {\tilde{Y}_0\bigl(\frac{m_n}{k}\bigl)}
      Y_M\Bigl(\frac{m_n}{k'}\Bigl) \Biggr)
  \Biggl( J_M\Bigl(\frac{m_n}{k'}\Bigl) 
    - \frac{J_1\bigl(\frac{m_n}{k}\bigl)}{Y_1\bigl(\frac{m_n}{k}\bigl)}
      Y_M\Bigl(\frac{m_n}{k'}\Bigl) \Biggr) = 0,
\label{eq:KKmass-U1gauginos}
\end{eqnarray}
where $\tilde{J}_0(m_n/k)$, $J_Y(m_n/k')$, $J_S(m_n/k')$ and $J_M(m_n/k')$ 
are defined by 
\begin{equation}
\begin{array}{lll} 
  \tilde{J}_0\left(\frac{m_n}{k}\right) 
  &\equiv& J_0\left(\frac{m_n}{k}\right)
    + \frac{g_B^2}{\tilde{g}_{0,1}^2}m_n J_1\left(\frac{m_n}{k}\right), \\
  J_Y\left(\frac{m_n}{k'}\right) 
  &\equiv& J_0\left(\frac{m_n}{k'}\right)
    + g_B^2 M_{\lambda,1} J_1\left(\frac{m_n}{k'}\right), \\
  J_S\left(\frac{m_n}{k'}\right) 
  &\equiv& J_0\left(\frac{m_n}{k'}\right)
    + g_B^2 M_{\lambda,S} J_1\left(\frac{m_n}{k'}\right), \\
  J_M\left(\frac{m_n}{k'}\right) 
  &\equiv& g_B^2 M_{\lambda,M} J_1\left(\frac{m_n}{k'}\right),
\end{array}
\end{equation}
and similarly for $\tilde{Y}_0(m_n/k)$, $Y_Y(m_n/k')$, $Y_S(m_n/k')$ 
and $Y_M(m_n/k')$.  Here $M_{\lambda}$'s are given by $M_{\lambda,1} 
= ((2/5)\zeta_C+(3/5)\zeta_R)\Lambda^{*2}/M_*$, $M_{\lambda,S} 
= ((3/5)\zeta_C+(2/5)\zeta_R)\Lambda^{*2}/M_*$ and $M_{\lambda,M} 
= (-(\sqrt{6}/5)\zeta_C+(\sqrt{6}/5)\zeta_R)\Lambda^{*2}/M_*$. 
In the supersymmetric limit, Eq.~(\ref{eq:KKmass-U1gauginos}) 
gives two decoupled KK towers for each of $\lambda^{321}_1$ 
and $\lambda^{\rm PS}_S$, reproducing the spectrum given in 
Eq.~(\ref{eq:spectrum-SO10-1}).  When supersymmetry is broken the two 
towers mix, but for small supersymmetry breaking the resulting tower 
can still be effectively described by the sum of the two independent 
towers for $\lambda^{321}_1$ and $\lambda^{\rm PS}_S$. The lightest 
state is almost purely $\lambda^{321}_1$ with the mass given as 
the lowest solution of Eq.~(\ref{eq:KKmass-32gauginos}) with $a=1$, 
and all the other states are heavier than $k'$ (the mixings are not 
negligible for the excited states).  With an increased strength for 
supersymmetry breaking, the mixing among the states becomes more 
important, giving, for example, a non-negligible effect on the mass 
of the lightest state.  For very large supersymmetry breaking, the 
lowest state is a Majorana fermion with the mass given by $\simeq 
(2k'/g_B^2)|M_{\lambda,1}/(M_{\lambda,S}M_{\lambda,1}-M_{\lambda,M}^2)| 
= 2((3/5)\zeta_C^{-1}+(2/5)\zeta_R^{-1})(M_*/g_B^2 \Lambda^{*2})k'$; 
the next state is a pseudo-Dirac fermion with the mass $\simeq 
g_1 k' (2/g_B^2 k)^{1/2} \approx 0.2\, k'$. 

The effects of supersymmetry breaking on the XY states are similar to 
those on the $SU(5)/321$ states in the model of~\cite{NST}.  Before 
supersymmetry breaking, the massless XY states consist of two Dirac 
fermions $\lambda'^{\rm XY}_X$ and $\lambda'^{\rm XY}_Q$, and four 
sets of real scalars $\sigma^{\rm XY}_X$, $\sigma^{\rm XY}_Q$, 
$A_{5,X}^{\rm XY}$ and $A_{5,Q}^{\rm XY}$, where the subscripts $X$ and 
$Q$ represent the $({\bf 3}, {\bf 2})_{-5/6} + ({\bf 3}^*, {\bf 2})_{5/6}$ 
and $({\bf 3}, {\bf 2})_{1/6} + ({\bf 3}^*, {\bf 2})_{-1/6}$ components 
under the 321 decomposition.  The masses for $\lambda'^{\rm XY}_X$ and 
$\lambda'^{\rm XY}_Q$ are generated through the operator
\begin{equation}
  S = \int\!d^4x \int_0^{\pi R}\!\!dy\,
    2 \delta(y-\pi R) \biggl[ e^{-2\pi kR} \int\!d^4\theta\, 
    \frac{\eta}{2 M_*} Z^\dagger \,
    {\rm Tr}[ {\cal P}[{\cal A}] {\cal P}[{\cal A}]] + {\rm h.c.} \biggr],
\label{eq:XYfermion-mass}
\end{equation}
where
\begin{equation}
  {\cal A} \equiv e^{-V}\!(\partial_y e^V) + (\partial_y e^V\!)\,e^{-V}
    - \sqrt{2}\, e^V \Sigma\, e^{-V} - \sqrt{2}\, e^{-V} \Sigma^\dagger e^V.
\label{eq:def-A}
\end{equation}
Here, the trace is taken over $SO(10)$ space and ${\cal P}[{\cal X}]$ 
is a projection operator: with ${\cal X}$ an adjoint of $SO(10)$, 
${\cal P}[{\cal X}]$ extracts the $({\bf 6}, {\bf 2}, {\bf 2})$ component 
of ${\cal X}$ under the decomposition to 422.  The coefficient $\eta$ is 
a dimensionless parameter.  Since the $X$ and $Q$ components are embedded 
in a single 422 multiplet $({\bf 6}, {\bf 2}, {\bf 2})$, the masses of 
their fermionic components are determined by  this single coefficient. The 
equation determining the $\lambda'^{\rm XY}_X$ and $\lambda'^{\rm XY}_Q$ 
masses is given by
\begin{equation}
  \frac{J_1\left(\frac{m_n}{k}\right)}
  {Y_1\left(\frac{m_n}{k}\right)}
  = \frac{J_1\left(\frac{m_n}{k'}\right)
    - g_B^2 M_{\lambda,X} J_0\left(\frac{m_n}{k'}\right)}
    {Y_1\left(\frac{m_n}{k'}\right)
    - g_B^2 M_{\lambda,X} Y_0\left(\frac{m_n}{k'}\right)},
\label{eq:KKmass-XYfermion}
\end{equation}
where $M_{\lambda,X} \equiv \eta \Lambda^2/M_*$.  Note that the masses 
for the $\lambda'^{\rm XY}_X$ and $\lambda'^{\rm XY}_Q$ towers are 
degenerate at tree level, although they split at loop level through 
the 321 gauge interactions (these splittings are finite and calculable 
as the XY towers are localized to the TeV brane while the breaking of 
422 resides at the Planck brane).  The masses for $\sigma^{\rm XY}_X$ 
and $\sigma^{\rm XY}_Q$ are generated through the operator
\begin{equation}
  S = \int\!d^4x \int_0^{\pi R}\!\!dy\,
    2 \delta(y-\pi R) \biggl[ -e^{-2\pi kR} \int\!d^4\theta\, 
    \frac{\rho}{4 M_*^2} Z^\dagger Z \,
    {\rm Tr}[ {\cal P}[{\cal A}] {\cal P}[{\cal A}]] \biggr],
\label{eq:XYscalar-mass}
\end{equation}
where the consistency of the effective theory requires $\rho$ to take 
the form $\rho = -8 g_B^2 |\eta|^2 \delta(0) + \rho'$, where $\rho'$ 
is a dimensionless parameter~\cite{NST}. The equation determining the 
$\sigma^{\rm XY}_X$ and $\sigma^{\rm XY}_Q$ masses is then given by
\begin{equation}
  \frac{J_1\left(\frac{m_n}{k}\right)}
  {Y_1\left(\frac{m_n}{k}\right)}
  = \frac{J_1\left(\frac{m_n}{k'}\right)
    - \frac{g_B^2 M_{\sigma,X}^2 k'}{m_n k} J_0\left(\frac{m_n}{k'}\right)}
    {Y_1\left(\frac{m_n}{k'}\right)
    - \frac{g_B^2 M_{\sigma,X}^2 k'}{m_n k} Y_0\left(\frac{m_n}{k'}\right)},
\label{eq:KKmass-XYscalar}
\end{equation}
where $M_{\sigma,X}^2 \equiv \rho' |\Lambda|^4/M_*^2$.  Again the masses 
for the $\sigma^{\rm XY}_X$ and $\sigma^{\rm XY}_Q$ towers are degenerate 
at tree level, although they split at loop level.  The masses of 
$A_{5,X}^{\rm XY}$ and $A_{5,Q}^{\rm XY}$ are not generated by 
the operators in Eqs.~(\ref{eq:XYfermion-mass},~\ref{eq:XYscalar-mass}). 
In fact, the 5D gauge invariance forbids any local operator giving these 
masses (in the 4D dual picture the zero modes of $A_{5,X}^{\rm XY}$ 
and $A_{5,Q}^{\rm XY}$ are pseudo-Goldstone bosons associated with the 
$SO(10) \rightarrow 422$ breaking at the TeV scale, which encodes the 
TeV-brane gauge breaking in 5D). These masses, however, are generated 
at loop level, picking up the effects of both Planck-brane and TeV-brane 
breakings.  The resulting masses are finite and approximately given by 
\begin{equation}
  m_{A_5^{\rm XY}}^2 \simeq \frac{g^2 C}{\pi^4}
    m_{\lambda'^{\rm XY}}^2 \ln\frac{\pi k'}{m_{\lambda'^{\rm XY}}},
\label{eq:A5XY-mass}
\end{equation}
where $g$ represents a 4D gauge coupling and $C$ the group theoretical 
factor.  The masses for $A_{5,X}^{\rm XY}$ and $A_{5,Q}^{\rm XY}$ are 
different, because they have the different 321 quantum numbers and 
thus different values of $C$s.  This difference, however, will be small, 
since the quantum numbers for $A_{5,X}^{\rm XY}$ and $A_{5,Q}^{\rm XY}$ 
are the same under $SU(3)_C$ and $SU(2)_L$ so that the mass difference 
only comes from the $U(1)_Y$ part, which is expected to give a mass 
splitting of $O(10\%)$. 

The Higgs spectrum is identical to the 321-321 model, except that the 
two $SU(5)$ multiplets, ${\bf 5}$ and ${\bf 5}^*$, of the 321-321 model 
are now unified into a single multiplet, ${\bf 10}$ of $SO(10)$.  As 
in~\cite{NST}, the unwanted zero modes in $H^c$ acquire TeV-scale masses 
through their tree-level couplings to the supersymmetry breaking on the TeV 
brane.  The Higgs spectrum of our model can be obtained from the expressions 
given in~\cite{NST} by setting $c_H^{} = c_{\bar{H}}$, $\eta_{H_D}^{} = 
\eta_{\bar{H}_D}^{}$ and $\eta_{H_T}^{} = \eta_{\bar{H}_T}^{}$.  Note, 
however, that the Planck-brane kinetic terms can still be different for 
the up-type and down-type Higgs fields, i.e. $z_H \neq z_{\bar{H}}$, due 
to the gauge breaking on the Planck brane.  This could be important for 
obtaining the correct electroweak symmetry breaking vacuum, depending on 
the details of the Higgs sector. 

It is useful here to consider a limit of small supersymmetry breaking 
$\Lambda \ll M_*$ and to see what the spectrum looks like. In this limit 
we can expand Eqs.~(\ref{eq:KKmass-32gauginos},~\ref{eq:KKmass-PSgauginos},%
~\ref{eq:KKmass-U1gauginos},~\ref{eq:KKmass-XYfermion},~%
\ref{eq:KKmass-XYscalar}) in powers of $\Lambda/M_*$.  We then find that 
the PS gauginos do not have any light mode with the mass smaller than 
$k'$, while the 321 gauginos, $\lambda^{321}_a$ ($a=1,2,3$), and the 
XY gauginos and scalars, $\lambda'^{\rm XY}_Z$ and $\sigma^{\rm XY}_Z$ 
($Z=X,Q$), do with the masses approximately given by
\begin{equation}
  m_{\lambda^{321}_a} = g_a^2 M_{\lambda,a}',
\label{eq:mass321-smallSB}
\end{equation}
\begin{equation}
  m_{\lambda'^{\rm XY}_Z} = 2 g_B^2 k M_{\lambda,X}',
\label{eq:massXYf-smallSB}
\end{equation}
\begin{equation}
  m_{\sigma^{\rm XY}_Z}^2 = 2 g_B^2 k M_{\sigma,X}'^2,
\label{eq:massXYs-smallSB}
\end{equation}
where $M_{\lambda,a}' \equiv M_{\lambda,a} e^{-\pi kR}$, $M_{\lambda,X}' 
\equiv M_{\lambda,X} e^{-\pi kR}$ and $M_{\sigma,X}' \equiv M_{\sigma,X} 
e^{-\pi kR}$ are parameters of order TeV, and $g_a \equiv (\pi R/g_B^2 
+ 1/\tilde{g}_{0,a}^2)^{-1/2}$ are the 4D gauge couplings.  Considering 
$g_a = O(1)$ and $g_B^2 k = O(\pi kR)$, we expect that the XY states 
are generically heavier than the 321 gauginos.  In fact, in the case 
of $M_{\lambda,a} \simeq M_{\lambda,X} \simeq M_{\sigma,X}/4\pi$, as 
suggested by naive dimensional analysis, the ratios of the masses 
are roughly given by $m_{\lambda^{321}_a} : m_{\lambda'^{\rm XY}_Z} : 
m_{\sigma^{\rm XY}_Z} \simeq 1 : \pi kR : \pi kR$.  The masses for 
$A_{5,Z}^{\rm XY}$ are given by Eq.~(\ref{eq:A5XY-mass}) so they could 
be somewhat lighter than $\lambda'^{\rm XY}_Z$ and $\sigma^{\rm XY}_Z$. 
Similarly in the Higgs sector, the colored-triplet states are generically 
heavier than the doublet states.  These little mass hierarchies among 
the TeV states arise because the wavefunctions for the exotic states 
are localized to the TeV brane, where supersymmetry breaking occurs, 
while those of the MSSM states are not.  This effect is therefore 
related to the model's successful prediction relating the low-energy 
gauge couplings, which crucially relies on the fact that all the exotic 
states are strongly localized to the TeV brane~\cite{NST}.  The spectrum 
for small supersymmetry breaking is depicted in Fig.~\ref{fig:spectrum}b 
for the gauge sector.

In the small supersymmetry breaking limit, the masses of the 321 gauginos, 
$\lambda^{321}_a$, are given by 
\begin{equation}
  M_1 = g_1^2\, \biggl( \frac{2}{5}\zeta_C + \frac{3}{5}\zeta_R \biggr) M',
\qquad
  M_2 = g_2^2\, \zeta_L M',
\qquad
  M_3 = g_3^2\, \zeta_C M',
\end{equation}
where $M_1 \equiv m_{\lambda^{321}_1}$, $M_2 \equiv m_{\lambda^{321}_2}$, 
and $M_3 \equiv m_{\lambda^{321}_3}$ are the bino, wino, and gluino masses, 
and $M' = e^{-\pi kR} \Lambda^{*2}/M_*$ is a parameter of order TeV.  
A particularly interesting case is where left-right symmetry is unbroken 
on the TeV brane. In this case $\zeta_L = \zeta_R$, so that we have 
a non-trivial relation among the 321 gaugino masses, which can be 
written as 
\begin{equation}
  \frac{M_1}{g_1^2}
    = \frac{2}{5}\frac{M_3}{g_3^2} + \frac{3}{5}\frac{M_2}{g_2^2}.
\label{eq:422-LR-rel}
\end{equation}
Note that, in contrast to high-scale supersymmetry breaking scenarios, 
this relation arises from the physics at an energy scale of order TeV 
as a threshold effect.%
\footnote{The relation Eq.~(\ref{eq:422-LR-rel}) is essentially determined 
by the symmetry of the theory --- 422 is preserved in the $G$ sector and 
its breaking comes only through the gauge kinetic terms of the gaugino 
fields.  Thus, Eq.~(\ref{eq:422-LR-rel}) holds accurately even for 
relatively small values of $M'_*/\pi k'$, i.e. it is not subject to errors 
of $O(\pi k'/M'_*)$ coming from unknown TeV-brane operators (note that 
$M_a$ here are the running masses and not the physical pole masses).}

We finally discuss the squark and slepton masses.  They are generated 
at one-loop level through the standard model gauge interactions. Because 
of the geometrical separation between supersymmetry breaking and the 
place where squarks and sleptons are located, the generated squark and 
slepton masses are finite and calculable in the effective field theory. 
Although the remaining gauge symmetry on the Planck brane after the 
orbifolding is 422, the squarks and sleptons interact effectively 
only with the 321 gauge multiplet at the scale where their masses are 
generated (due to the spontaneous breaking of 422 at a high scale). 
This means that the squark and slepton masses in the present model 
are given by
\begin{equation}
  m_{\tilde{f}}^2 = \frac{1}{2\pi^2}\! 
    \sum_{a=1,2,3}\! {\cal F}^{\delta_{a1}} C_a^{\tilde{f}}\, {\cal I}_a,
\label{sfermion-model1}
\end{equation}
where $\tilde{f} = \tilde{q}, \tilde{u}, \tilde{d}, \tilde{l}, \tilde{e}$ 
represents the MSSM squarks and sleptons, and the $C_a^{\tilde{f}}$ are 
the group theoretical factors given by $(C_1^{\tilde{f}}, C_2^{\tilde{f}}, 
C_3^{\tilde{f}}) = (1/60,3/4,4/3)$, $(4/15,0,4/3)$, $(1/15,0,4/3)$, 
$(3/20,3/4,0)$ and $(3/5,0,0)$ for $\tilde{f} = \tilde{q}, \tilde{u}, 
\tilde{d}, \tilde{l}$ and $\tilde{e}$, respectively.  The functions 
${\cal I}_a$ are defined in Eq.~(21) of~\cite{Nomura:2003qb}, where 
we have to use three different gaugino mass parameters $M_{\lambda,a}$ 
because of the non-universal gaugino masses.%
\footnote{The expressions for squark and slepton masses, 
Eq.~(\ref{sfermion-model1}), are subject to errors of $O(\pi k'/M'_*)$ 
arising from the TeV-brane localized 321 gauge kinetic terms (only errors 
of $O((\pi k'/M'_*)^2)$ are mentioned in~\cite{Nomura:2003qb}).}
The quantity ${\cal F}$, which represents a mixing effect between the 
$U(1)_Y$ and $U(1)_\chi$ gaugino towers, is a function of $M_{\lambda,1}$, 
$M_{\lambda,S}$ and $M_{\lambda,M}$ and takes a value of order 1.  Since 
the mixing effect vanishes for small supersymmetry breaking, ${\cal F}$ 
approaches to $1$ for small $M_\lambda$'s.  Note that because the squark 
and slepton masses are generated through the gauge interactions, they are 
flavor universal and the supersymmetric flavor problem is absent.  Small 
mass splittings among different generations arise through the Yukawa 
couplings at two loop orders, making the masses for the third generation 
squarks and sleptons slightly lower than those for the fist two 
generation ones.  However, they do not generate flavor changing 
neutral currents at a dangerous level. 

The spectrum described above provides a rich phenomenology.  For 
example, it gives a variety of possibilities for the next-to-lightest 
supersymmetric particle (NLSP) (the lightest supersymmetric particle is 
the gravitino with the mass $\sim k'^2/M_{Pl} \sim 0.01-0.1~{\rm eV}$). 
Possible patterns for the superparticle spectrum are similar to those 
discussed in~\cite{NST}.  The model also predicts relatively light 
GUT particles $A_{5,X}^{\rm XY}$ and $A_{5,Q}^{\rm XY}$, transforming 
as $({\bf 3}, {\bf 2})_{-5/6} + ({\bf 3}^*, {\bf 2})_{5/6}$ and 
$({\bf 3}, {\bf 2})_{1/6} + ({\bf 3}^*, {\bf 2})_{-1/6}$ under 321 
respectively.  The masses for $A_{5,X}^{\rm XY}$ and $A_{5,Q}^{\rm XY}$ 
are close but expected to have a relative splitting of $O(10\%)$. The 
lighter one will presumably be $A_{5,Q}^{\rm XY}$, which is stable 
for collider purposes due to the conservation of $SU(3)_C$ charges 
and the location of fields~\cite{NST}.  Once $A_{5,Q}^{\rm XY}$ is 
produced, it hadronizes to either of four fermionic mesons $\hat{T}^0$, 
$\hat{T}^{'0}$, $\hat{T}^{+}$, $\hat{T}^{'+}$ (and their anti-particles), 
depending on whether it picks up an up or down quark or anti-quark. 
All these states are sufficiently long-lived, so that the charged 
ones would be detectable through highly ionizing tracks. For strong 
supersymmetry breaking (large $\Lambda$), the gauginos of the PS multiplet 
$\lambda'^{\rm PS}_U$, $\lambda'^{\rm PS}_E$ and $\lambda'^{\rm PS}_S$, 
transforming as $({\bf 3}, {\bf 1})_{2/3} + ({\bf 3}^*, {\bf 1})_{-2/3}$, 
$({\bf 1}, {\bf 1})_1 + ({\bf 1}, {\bf 1})_{-1}$ and $({\bf 1}, {\bf 1})_0$, 
also become light (see Fig.~\ref{fig:spectrum}c for the overall spectrum 
of the gauge sector for large supersymmetry breaking).  Some of these 
states could also be stable and seen at colliders.

\section{Model without TeV-Brane Symmetry Breaking}
\label{sec:model-2}

In this section we construct a model in which the bulk gauge symmetry 
is not reduced at the TeV brane.  The construction closely follows that 
of the previous section.  The model presented here can be viewed as an 
extension of the $SU(5)$ model of~\cite{Goldberger:2002pc} to a larger 
unified gauge group. 

The model is again formulated in the 5D warped spacetime compactified 
on $S^1/Z_2$ ($0 \leq y \leq \pi R$), with the metric given by 
Eq.~(\ref{eq:metric}).  The parameters $M_5$, $M_*$, $k$ and $R$ take 
similar values.  The gauge group in the bulk is taken to be $SO(10)$, 
with the boundary conditions given by
\begin{equation}
  \pmatrix{V \cr \Sigma}(x^\mu,-y) 
  = \pmatrix{P V P^{-1} \cr -P \Sigma P^{-1}}(x^\mu,y), 
\qquad
  \pmatrix{V \cr \Sigma}(x^\mu,-y') 
  = \pmatrix{V \cr -\Sigma}(x^\mu,y'), 
\label{eq:bc-g-2}
\end{equation}
where $y' = y-\pi R$. The matrix $P$ is the same as before: $P = \sigma_0 
\otimes {\rm diag}(1,1,1,-1,-1)$.  This reduces the gauge group at the 
Planck brane to 422, but leaves the gauge group at the TeV brane to 
be $SO(10)$.  The resulting zero modes are only the $({\bf 15}, {\bf 1}, 
{\bf 1}) + ({\bf 1}, {\bf 3}, {\bf 1}) + ({\bf 1}, {\bf 1}, {\bf 3})$ 
component of $V$ (the 422 $N=1$ gauge supermultiplet).  The excited 
KK states all have masses of order $k'$ or larger. 

The Higgs fields are introduced in the bulk as a hypermultiplet 
transforming as ${\bf 10}$ of $SO(10)$, with the boundary conditions 
given by
\begin{equation}
  \pmatrix{H \cr H^c}(x^\mu,-y) 
  = \pmatrix{-P H \cr P H^c}(x^\mu,y), 
\qquad
  \pmatrix{H \cr H^c}(x^\mu,-y') 
  = \pmatrix{H \cr -H^c}(x^\mu,y').
\label{eq:bc-h-2}
\end{equation}
This leaves the $({\bf 1}, {\bf 2}, {\bf 2})$ component of $H$ as 
zero modes.  All the other modes have masses of order $k'$ or larger.%
\footnote{We can alternatively put the Higgs field on the Planck brane 
in the $({\bf 1}, {\bf 2}, {\bf 2})$ representation of 422, or in the 
bulk but with the boundary conditions of Eq.~(\ref{eq:bc-h-2}) with 
an extra minus sign in the right-hand side of the second equation 
(cf.~footnote~\ref{ft:Higgs}).  The latter case gives two triplet zero 
modes from $H^c$ and four relatively light doublet modes from $H$ and 
$H^c$ (two for each) in the supersymmetric limit (the doublet states 
are even exponentially lighter than $k'$ for $c_H > 1/2$).  A realistic 
model is then obtained by introducing a mass term of the form 
$\int\!d^2\theta H^c H^c + {\rm h.c.}$ on the TeV brane.  The MSSM 
prediction for gauge coupling unification is preserved in these cases, 
too (for $c_H \geq 1/2$ in the case of the bulk Higgs).}

The matter and the 422 gauge breaking sectors are identical to those 
in the model of section~\ref{sec:model-1}.  The quark and lepton 
chiral superfields are introduced on the Planck brane in the 
$\Psi({\bf 4}, {\bf 2}, {\bf 1}) + \bar{\Psi}({\bf 4}^*, {\bf 1}, 
{\bf 2})$ representation of 422 for each generation.%
\footnote{In the present model, (some of) matter fields could be 
introduced in the bulk as hypermultiplets transforming as ${\bf 16}$ 
of $SO(10)$ (a hypermultiplet transforming as ${\bf 16}$ of $SO(10)$ 
yields either $\Psi$ or $\bar{\Psi}$ as a zero mode, depending on the 
boundary conditions). The prediction for the low-energy gauge couplings 
is the same as that in the brane matter case if the bulk mass parameters, 
$c_M$, for matter fields take the values $c_M \geq 1/2$.}
The 422 gauge breaking is introduced on the Planck brane. The simplest 
possibility is the Higgs breaking $\langle \chi \rangle = \langle 
\bar{\chi} \rangle = O(k)$, where $\chi$ and $\bar{\chi}$ transform 
as $({\bf 4}, {\bf 1}, {\bf 2})$ and $({\bf 4}^*, {\bf 1}, {\bf 2})$ 
under 422, respectively. We introduce the Planck brane couplings 
Eq.~(\ref{eq:yukawa-422}), which give the Yukawa couplings for the 
quarks and leptons as well as Majorana masses for the right-handed 
neutrinos.  With suitable non-renormalizable operators realistic quark 
and lepton masses and mixings are reproduced.  For $\langle \chi \rangle, 
\langle \bar{\chi} \rangle \simeq k \simlt M_*$, the Yukawa couplings 
for the third generation fermions are approximately unified at the 
scale $k$, giving a successful low-energy prediction for $m_b/m_\tau$. 
It also leads to $\tan\beta \approx 50$.  Small masses for the observed 
neutrinos are naturally obtained through the see-saw mechanism.  The 
model possesses a $U(1)_R$ symmetry: $V(0), \Sigma(0), H(0), H^c(2), 
\Psi(1), \bar{\Psi}(1), \chi(0), \bar{\chi}(0)$ (broken to $Z_{2,R}$ 
at the TeV brane through supersymmetry breaking), which prevents 
potentially dangerous operators such as $\delta(y)\int\!d^2\theta 
H_D^2 + {\rm h.c.}$.  There is not a proton decay problem for 
Planck-brane localized matter.

The 422 breaking at the Planck brane gives masses for the 422/321 
component of $V$.  The resulting spectrum for the gauge sector is 
then given by Eq.~(\ref{eq:KKmass-321}) for the 321 component and 
by Eq.~(\ref{eq:KKmass-PS}) for the $SO(10)/321$ component (with 
the comment in footnote~\ref{ft:comment} applying to 422/321). 
The spectrum, therefore, can be summarized as
\begin{equation}
  \left\{ \begin{array}{ll} 
    V^{321} : & m_0 = 0, \\
    \{ V^{321}, \Sigma^{321} \} + \{ V^{\rm GUT}, \Sigma^{\rm GUT} \}: 
      & m_n \simeq (n-\frac{1}{4})\pi k',
  \end{array} \right.
\label{eq:spectrum-SO10-2}
\end{equation}
where $n = 1,2,\cdots$.  Here, the superscript GUT represents 
the $SO(10)/321$ component, whose transformation properties are 
$({\bf 3}, {\bf 2})_{-5/6} + ({\bf 3}^*, {\bf 2})_{5/6} + ({\bf 3}, 
{\bf 2})_{1/6} + ({\bf 3}^*, {\bf 2})_{-1/6} + ({\bf 3}, {\bf 1})_{2/3} 
+ ({\bf 3}^*, {\bf 1})_{-2/3} + ({\bf 1}, {\bf 1})_1 + ({\bf 1}, 
{\bf 1})_{-1} + ({\bf 1}, {\bf 1})_0$ under the 321 decomposition. 
Note that the spectrum for the excited KK towers are approximately 
$SO(10)$ symmetric.  This is because in the 4D dual picture the 
global $SO(10)$ symmetry of $G$ is not dynamically broken so that 
the composite states of $G$, which are identified as KK towers 
in 5D, are approximately $SO(10)$ symmetric. 

As in the model of section~\ref{sec:model-1}, the MSSM prediction for 
low-energy gauge couplings is preserved.  Specifically, low-energy 321 
gauge couplings are given by setting $(T_1,T_2,T_3)(V_{++}) = (0,2,3)$, 
$(T_1,T_2,T_3)(V_{-+}) = (8,6,5)$ and $(T_1,T_2,T_3)(V_{+-}) = 
(T_1,T_2,T_3)(V_{--}) = (0,0,0)$ in Eq.~(9) of Ref.~\cite{Goldberger:2002pc} 
and adding the Higgs contribution, which is identical to that of the 
$SU(5)$ model of~\cite{Goldberger:2002pc} with $c_H = c_{H({\bf 5})} = 
c_{\bar{H}({\bf 5}^*)}$.  Here, $c_H$ is the dimensionless bulk 
mass parameter for the Higgs multiplet, $\{ H, H^c \}$.  Therefore, 
for $c_H \geq 1/2$ we find that the prediction for low-energy 321 
gauge couplings $g_a$ is given by Eq.~(\ref{eq:gc-low-1}) with 
Eq.~(\ref{eq:gc-low-2}), reproducing the successful MSSM prediction. 

Supersymmetry breaking is introduced on the TeV brane through the 
potential Eq.~(\ref{eq:Z-TeV}).  This gives the vacuum expectation 
value $\langle Z \rangle = - e^{-\pi kR} \Lambda^{*2} \theta^2$, 
breaking supersymmetry and the $U(1)_R$ symmetry.  The breakings 
are transmitted to various fields through the operator 
\begin{equation}
  S = \int\!d^4x \int_0^{\pi R}\!\!dy\,
    2 \delta(y-\pi R) \biggl[ 
      -\int\!d^2\theta\, \frac{\zeta}{2 M_*} Z \, 
      {\rm Tr}[ {\cal W}^\alpha {\cal W}_\alpha ] 
      + {\rm h.c.} \biggr],
\label{eq:univ-gaugino-mass}
\end{equation}
where the trace is taken over $SO(10)$ space. Since the TeV brane 
respects full $SO(10)$, the coefficient $\zeta$ is universal for the 
entire gauge components.  The above operator gives masses for the 
321 gaugino zero modes and modifies the spectrum for the gaugino 
towers of both 321 and $SO(10)/321$ components.  The masses for the 
321 and $SO(10)/321$ gaugino towers are given, respectively, by 
Eq.~(\ref{eq:KKmass-32gauginos}) and Eq.~(\ref{eq:KKmass-PSgauginos}), 
but now the gaugino mass parameters $M_\lambda$'s take a universal 
value: $M_{\lambda,a} = M_{\lambda,A} = \zeta \Lambda^{*2}/M_* 
\equiv M_\lambda$.  The values of the Planck-brane gauge couplings 
renormalized at the TeV scale are the same as in the previous model: 
$1/\tilde{g}_{0,a}^2 \simeq \Delta^a/8\pi^2$ should be used in 
Eq.~(\ref{eq:KKmass-32gauginos}). 

For the Higgs sector, the only light states are those of the two 
MSSM Higgs doublets.  They obtain supersymmetry-breaking as well as 
supersymmetry-preserving masses through the couplings to the $Z$ 
field on the TeV brane.  The resulting masses are identical to 
those in the model of section~\ref{sec:model-1}, whose explicit 
expressions are given in Ref.~\cite{NST}.

Squark and slepton masses are generated at one-loop level through 
gauge interactions.  Despite the 422 symmetry after the orbifolding, 
the squark and slepton masses come almost entirely from the 321 gauge 
loops because 422 is broken at a high scale. The resulting masses 
are finite and calculable, due to the spatial separation between 
supersymmetry breaking and the matter location, and are given by
\begin{equation}
  m_{\tilde{f}}^2 = 
    \frac{1}{2\pi^2}\! \sum_{a=1,2,3}\! C_a^{\tilde{f}}\, {\cal I}_a,
\end{equation}
where $\tilde{f} = \tilde{q}, \tilde{u}, \tilde{d}, \tilde{l}, 
\tilde{e}$, and $C_a^{\tilde{f}}$ are the group theoretical factors 
given below Eq.~(\ref{sfermion-model1}).  The functions ${\cal I}_a$ 
are defined in~\cite{Nomura:2003qb}, where we have to use the 
gaugino mass parameter $M_\lambda$ for all $a=1,2,3$.  Because of 
the universality of the operator Eq.~(\ref{eq:univ-gaugino-mass}), 
the effect of the $U(1)_Y$-$U(1)_\chi$ mixing is negligible.  The 
supersymmetric flavor problem is absent because the squark and slepton 
masses are flavor universal (up to small higher-order corrections 
from the Yukawa couplings). 

The phenomenology of the present model is similar to that of the 
model in~\cite{Goldberger:2002pc}.  In fact, the masses for all the 
superparticles as well as the KK towers are all determined in terms 
of only two parameters $M_\lambda/k$ and $k'$, up to parameters associated 
with the Higgs sector~\cite{Nomura:2003qb}.  The main difference is 
that the $SU(5)/321$ gauge multiplet of~\cite{Goldberger:2002pc} is now 
replaced by the $SO(10)/321$ multiplet.  Therefore, the light exotics now 
consist of five components transforming as $({\bf 3}, {\bf 2})_{-5/6} 
+ ({\bf 3}^*, {\bf 2})_{5/6}$,  $({\bf 3}, {\bf 2})_{1/6} + ({\bf 3}^*, 
{\bf 2})_{-1/6}$, $({\bf 3}, {\bf 1})_{2/3} + ({\bf 3}^*, {\bf 1})_{-2/3}$, 
$({\bf 1}, {\bf 1})_1 + ({\bf 1}, {\bf 1})_{-1}$, and $({\bf 1}, 
{\bf 1})_0$ under 321. In particular, in the case of large supersymmetry 
breaking the gauginos for all these components become light, thus 
providing the possibility of testing the underlying enlarged group 
structure of the model at the LHC.%
\footnote{In the present model, the $Z_2$ GUT parity 
of~\cite{Goldberger:2002pc} is extended to two $Z_2$ parities: 
one which acts non-trivially on the $SO(10)/422$ component of the 
gauge multiplet and the colored Higgs states and the other which 
acts non-trivially on the ``$SO(10)/(SU(5) \times U(1)_\chi)$'' 
component of the gauge multiplets, i.e. the $U$ and $E$ of PS and 
$Q$ of XY.  These symmetries ensure the quasi-stability for the 
lightest of $\lambda'^{\rm XY}_X$ and $\lambda'^{\rm XY}_Q$ and of 
$\lambda'^{\rm XY}_Q$, $\lambda'^{\rm PS}_U$ and $\lambda'^{\rm PS}_E$, 
respectively.}

\section{Conclusions}
\label{sec:concl}

In this paper we have seen that the Pati-Salam (422) gauge group fits 
extremely naturally into the framework of warped unification.  The two 
models we have constructed both have a bulk $SO(10)$ gauge symmetry 
broken by boundary conditions to 422 on the Planck brane, corresponding 
in the 4D picture to a 422 gauge theory with an approximate global 
$SO(10)$ symmetry.   They differ in whether the full $SO(10)$ is 
realized on the TeV brane, or equivalently, in whether or not the 
global $SO(10)$ is spontaneously broken in the infrared by the strong 
dynamics of the conformal sector. 

These models incorporate matter unification in a very economical way, 
with a full generation of quarks and leptons transforming as $({\bf 4}, 
{\bf 2}, {\bf 1}) + ({\bf 4}^*, {\bf 1}, {\bf 2})$ under 422. They also 
require only a very simple Higgs sector for breaking of the 422 gauge 
symmetry and easily accommodate the see-saw mechanism for neutrino masses, 
realistic fermion masses, and bottom-tau unification.  At the same time 
they explain the successful unification of gauge couplings in the MSSM, 
something that Pati-Salam unification by itself does not do.  The same 
prediction relating the low energy couplings given in the MSSM applies 
here to good approximation under the assumption that the gauge interactions 
grow strong in the ultraviolet.  The $SO(10)$ symmetry plays an important 
role in this prediction; in the 4D description, it ensures that the 
contribution to the evolution of the gauge couplings from the conformal 
sector is universal.%
\footnote{It is worth noting that the $SO(10)$-symmetric conformal 
sector can be replaced by extra vector-like states, such as 
$({\bf 4}, {\bf 2}, {\bf 1}) + ({\bf 4}^*, {\bf 2}, {\bf 1})$ and 
$({\bf 4}, {\bf 1}, {\bf 2}) + ({\bf 4}^*, {\bf 1}, {\bf 2})$ with 
TeV-scale masses, without spoiling many of the desired features of 
the model.  This provides a class of purely 4D 422 theories with the 
successful gauge coupling prediction arising from strong coupling 
in the ultraviolet~\cite{LNS}.}

The phenomenology of these models is quite different from conventional 
supersymmetric unification, and different as well from models of warped 
unification built on $SU(5)$ symmetry. In both models a rich array 
of exotic particles appears near the TeV scale: these include 
supermultiplets with 321 quantum numbers $({\bf 3}, {\bf 2})_{-5/6} 
+ ({\bf 3}^*, {\bf 2})_{5/6}$ as in warped $SU(5)$, but also states 
with quantum numbers $({\bf 3}, {\bf 2})_{1/6} + ({\bf 3}^*, 
{\bf 2})_{-1/6}$, $({\bf 3}, {\bf 1})_{2/3}+({\bf 3}^*, {\bf 1})_{-2/3}$, 
and color-neutral states transforming as $({\bf 1}, {\bf 1})_{1} + 
({\bf 1}, {\bf 1})_{-1}+({\bf 1}, {\bf 1})_{0}$. In the model with 
$SO(10)$ broken to 422 on the TeV brane, some of these states are 
massless in the supersymmetric limit.  The prospects for producing 
these particles at future colliders such as the LHC depend on 
the scale $k'$, the strength of supersymmetry breaking on the TeV 
brane, and in the model with TeV-brane symmetry breaking, on the 
free parameters that determine how strongly the pseudo-Goldstone 
multiplet feels the supersymmetry breaking ($\eta$ and $\rho$ of 
Eqs.~(\ref{eq:XYfermion-mass},~\ref{eq:XYscalar-mass})).

The spectrum of MSSM superparticles differs between the two models. In 
the model with TeV-brane symmetry breaking, the gaugino mass terms on 
the TeV brane are non-universal, although there is the possibility of 
one relation among the 321 gaugino masses if left-right symmetry is 
unbroken on the TeV brane.  One interesting point is that in the TeV-brane 
symmetry breaking model, there is generally a mixing between the bino and 
the gaugino associated with the other $U(1)$ factor  contained in $SU(4)_C 
\times SU(2)_R$.  The effect of this mixing is small for weak supersymmetry 
breaking, but becomes significant as the supersymmetry breaking is increased. 
Because of the non-universality in the gaugino masses, there is a broad 
range of possibilities for what the gaugino and scalar spectrum will look 
like, and in particular, there are many possibilities for what the NLSP 
will be, as discussed in~\cite{NST}.  The model without TeV-brane symmetry 
breaking, on the other hand, has a more constrained spectrum due to the 
universal gaugino mass terms on the TeV brane.  The spectrum of MSSM 
particles has similar features to that of~\cite{Goldberger:2002pc,%
Nomura:2003qb} in this case, but it should be stressed again that 
the spectrum of exotic particles is different.

\section*{Acknowledgment}

The work of Y.N. was supported in part by the Director, Office 
of Science, Office of High Energy and Nuclear Physics, of the U.S. 
Department of Energy under Contract DE-AC03-76SF00098.

\end{document}